\documentclass[pre,a4paper,onecolumn]{revtex4}

\usepackage{graphicx}
\usepackage{amsmath}
\usepackage{bbm}
\usepackage{psfrag}
\usepackage{latexsym}

\renewcommand{\d}{\text{d}}

\begin{document} 

\title{Random matrix model for QCD$_\mathbf{3}$ staggered fermions}

\author{P. Bialas$^{1,2}$}{\thanks{pbialas@th.if.uj.edu.pl}
  \author{Z. Burda$^{1,2}$}{\thanks{zdzislaw.burda@uj.edu.pl}
    \author{B. Petersson$^{3,4}$}{\thanks{bengt@physik.hu-berlin.de}
      \affiliation{ $^1$Marian Smoluchowski Institute of Physics,
        Jagiellonian University, Reymonta 4, 30-059 Krak\'ow, Poland
        \\ $^2$Mark Kac Complex Systems Research Centre, Jagiellonian
        University, Reymonta 4, 30-059 Krak\'ow, Poland \\ $^3$
        Faculty of Physics, University of Bielefeld, P.O. Box 10 01
        21, D-33501 Bielefeld, Germany \\ $^4$ Institute of Physics,
        Humboldt University, Newtonstr. 15, D-12489 Berlin, Germany }
\begin{abstract}
  We show that the lowest part of the eigenvalue density of the
  staggered fermion operator in lattice QCD$_3$ at small lattice coupling constant $\beta$  has exactly the same shape as in QCD$_4$. This observation
  is quite surprising, since universal properties of the QCD$_3$ Dirac
  operator are expected to be described by a non-chiral matrix
  model. We show that this effect is related to the specific nature of
  the staggered fermion discretization and that the eigenvalue density
  evolves towards the non-chiral random matrix prediction when $\beta$
  is increased and the continuum limit is approached.  We propose a
  two-matrix model with one free parameter which interpolates between
  the two limits and very well mimics the pattern of evolution with
  $\beta$ of the
  eigenvalue density of the staggered fermion operator in QCD$_3$.
\end{abstract}
\maketitle

\section*{Introduction}

One can argue, referring to the universality \cite{sv}, that the low
energy properties of the QCD$_4$ Dirac operator in the
$\epsilon$-regime are described by the chiral random matrix model
\cite{vz4}. In particular, the microscopic eigenvalue density of the
Dirac operator, which is obtained from the eigenvalue density
\begin{equation}
\rho(\lambda) = \left\langle \sum_i \delta(\lambda-\lambda_i) \right\rangle
\end{equation}
by blowing it up at $\lambda=0$, is expected to have the same
universal shape as in the random matrix. More precisely, the
microscopic eigenvalue density is defined as
\begin{equation} 
\rho_{*}(\lambda) = \lim_{N\rightarrow \infty} \frac{1}{N\Sigma} \; 
\rho\left(\frac{\lambda}{N\Sigma}\right) \ ,
\label{micro}
\end{equation}
where $\Sigma = \pi \rho(0)$ and $N$ is the number of eigenvalues'
pairs of the underlying discretized (regularized) Dirac operator, $N$
is related to the physical volume $V \sim N a^D$ where $a$ is an
UV-cut-off.  The microscopic density $\rho_*(\lambda)$ can be
determined analytically using the random matrix model. For the trivial
topological sector and for $N_f=0$ flavors it reads \cite{vz4}
\begin{equation} 
\rho_{4*}(\lambda) = \frac{\lambda}{2} \left( J^2_0(\lambda)  
+ J^2_1(\lambda) \right) \ . 
\label{rho4} 
\end{equation}
This eigenvalue density has been compared to numerical data from
quenched Monte-Carlo simulations of lattice QCD with staggered
fermions \cite{sf}.  One could indeed see a very good agreement
between the random matrix prediction and lattice data \cite{dhk}.
It has later been shown that with an improved staggered action also the non trivial topological sectors are well described \cite{follana}. 

If one applies the same arguments to QCD$_3$ one is led to a non-chiral
random matrix model and to the following microscopic density \cite{vz3} 
\begin{equation} 
\rho_{3*}(\lambda) = \frac{1}{\pi} \ .
\label{rho3} 
\end{equation}
Also this prediction can be easily tested numerically by comparing it
to lattice QCD$_3$ data. Actually some tests have been done in the
past but they were not conclusive \cite{dhkm}. In this work we repeat
quenched simulations of lattice QCD$_3$ to determine the eigenvalue
density of the staggered fermion operator. The spectrum of the lowest
positive eigenvalues of the staggered fermion operator for quenched
QCD$_3$ with $\beta=6.0$ is shown in figure \ref{puzzle}. As one can
see the result is quite surprising. The QCD$_3$ data matches the
chiral prediction (\ref{rho4}) and not the non-chiral one (\ref{rho3})
anticipated in this case.
\begin{figure} 
\begin{center} 
\includegraphics[width=12cm]{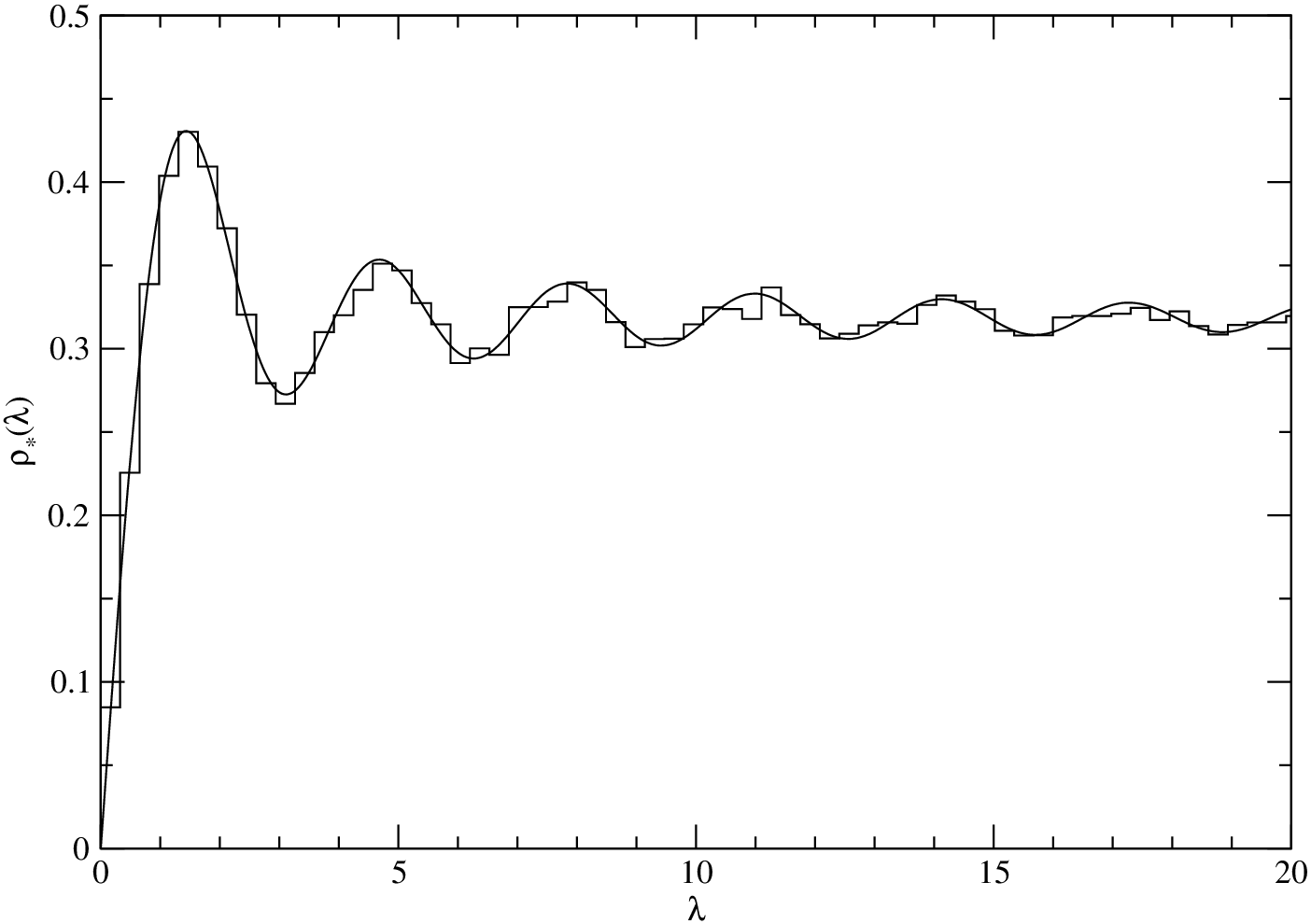}
\end{center} 
\caption{\label{puzzle} Comparison of the lowest part 
of the eigenvalue distribution of the staggered fermion 
operator obtained in Monte-Carlo simulations of quenched QCD$_3$ 
with the $SU(3)$ Wilson action for $\beta=6$ on a $14^3$ 
lattice to the chiral random matrix prediction
(blue dotted line) and the corresponding non-chiral random matrix prediction (green solid line). 
The data fits perfectly to the chiral random matrix density.} 
\end{figure}
This finding raises the following questions for discussion.  Why a
three dimensional theory which has no chiral symmetry shows a pattern
characteristic for a chiral theory? Is the effect related to the
regularization scheme? If yes, will it disappear when the continuum
limit is approached? In QCD$_4$ where the chiral symmetry plays an
important role, the pattern will stay intact when $\beta$ is increased
and the continuum limit is approached while in QCD$_3$ the shape of
the lowest part of the spectrum should gradually evolve from
(\ref{rho4}) to (\ref{rho3}). These questions will be addressed in the
remaining part of the paper. The paper is organized as follows.  In
the next section we recall the matrix models used to describe the
universal properties of the lowest part of the Dirac operator spectrum
in four and three dimensions. We emphasize the difference between
them.  Then we discuss staggered fermions and introduce a new matrix
model with one free parameter to imitate the evolution with $\beta$ of
the staggered fermion operator spectra in QCD$_3$. It describes a
gradual disappearance of a hard edge at the origin of the spectrum
when $\beta$ goes to infinity. We compare the low energy part of the
eigenvalue distribution of quenched QCD$_3$ and of the random matrix
model. We conclude the paper with a short summary.

\section*{Random matrix model}

The relation of the low energy spectrum of the Dirac operator in four
dimensions to the chiral random matrix model has been extensively
discussed in the literature so here we will restrict ourselves to a
minimal presentation which is concentrated on the difference between
the chiral and the non-chiral case. The interested reader is referred
to reviews \cite{v,vw} and the references therein. In four dimensions
the Euclidean Dirac operator assumes in the chiral representation the
following form
\begin{equation} 
{\cal D} = \left( \begin{array}{cc} 0 & i{\cal W}^\dagger \\ 
i{\cal W} & 0 \end{array}\right) 
\label{D} 
\end{equation} 
where 
\begin{equation} 
{\cal W} = {\cal W}_4 \mathbbm{1}  + i \sum_{k=1}^3 {\cal W}_k \sigma_k  \ ,
\label{W4} 
\end{equation} 
$\mathbbm{1}$ is a $2\times 2$ unit matrix, $\sigma_k$, $k=1,2,3$ are
Pauli matrices. The operators ${\cal W}_\mu = -i \partial_\mu +
A_\mu(x)$, $\mu=1,2,3,4$ are hermitian. In addition to the space-time
indices they contain also $SU(3)$ group indices which are suppressed
in our notation.

The corresponding random matrix model is obtained by 
replacing  ${\cal W}$ by an $N\times N$ complex random matrix $D$ that
inherits the block structure and basic symmetries of the operator ${\cal D}$ 
\begin{equation} 
D_4 = \left( \begin{array}{cc} 0 & iW^\dagger \\ 
iW & 0 \end{array}\right) \ .
\label{D4}
\end{equation} 
The simplest candidate for the matrix $W$ is a Gaussian random matrix generated with the probability measure
\begin{equation} 
\d\mu_4(W) = \d W e^{-N\Sigma^2 \; {\rm Tr} W^\dagger W} \ , 
\label{mu4} 
\end{equation} 
where $\d W = \sum_{ij} \d {\rm Re} W_{ij} \; \d {\rm Im} W_{ij}$.  This
choice may look arbitrary at first sight but luckily it is known that
the microscopic properties of the eigenvalue density of random
matrices exhibit a large degree of universality which means that the
microscopic density does not change for a large class of probability
measures as long as they do not change the symmetry or the nature of
the noise of the random matrix or introduce strong long range
correlations. For example one can rigorously prove that the
microscopic eigenvalue density of any random matrix generated with a
probability measure $\d W \exp -N {\rm Tr} V(W^\dagger W)$ for a
polynomial potential $V$ is exactly the same as for the Gaussian
measure (\ref{mu4}). The microscopic properties can be however
modified by adding non-analytic terms to the potential like for
instance logarithmic ones. Actually such logarithmic terms come
naturally into play in QCD if one integrates out fermionic degrees of
freedom. Here we will however restrict ourselves only to the quenched
approximation where such terms are absent. In this case the Gaussian
measure is the simplest and the best candidate. The behavior of the
lowest part of the spectrum can be also modified when one replaces a
square matrix $W$ by a rectangular $N_+ \times N_-$ one, where
$N_++N_-=2N$. Such a modification introduces $\nu =N_+-N_-$
right-handed zero modes to the matrix $D$, if $N_+>N_-$, or
$\nu=N_--N_+$ left-handed ones, if $N_->N_+$, imitating different
topological sectors. In the trivial topological sector, which we
consider here, $W$ can be viewed as a square Gaussian complex matrix
(\ref{mu4}). The microscopic eigenvalue distribution of the matrix $D$
(\ref{D4}) can be calculated analytically. The calculations \cite{vz4}
give the expression mentioned above in eq. (\ref{rho4}).

Consider now three dimensions. In this case there is no chiral
symmetry as we will see below but one can define a two fermion family
model which imitates this symmetry. The spinors have two components in
the fundamental representation. There are two independent
nonequivalent sets of gamma matrices $\gamma_k=\sigma_k$ and
$\gamma'_k=-\sigma_k$, where as before $\sigma_k$ are Pauli
matrices. One can associate a fermion family with each representation
and consider a common theory for the two families.  A Lagrangian for
such a theory can be concisely written in terms of four component
spinors \cite{p}. It takes a very similar form to the four dimensional
Lagrangian. In particular it has a U(2) symmetry, which is broken by a
mass term to a $U(1)\otimes U(1)$ symmetry. We thus have a situation
similar to the chiral symmetry in four dimensions.  The main
difference is that the term with $\gamma_4$ is absent in the
Lagrangian. The fermionic operator takes the form
\begin{equation} 
{\cal W} = i \sum_{k=1}^3 {\cal W}_k \sigma_k \ .
\label{W3} 
\end{equation} 
So in three dimension the operator ${\cal W}$ is 
anti-hermitian ${\cal W}=-{\cal W}^\dagger$. Thus if one constructs a matrix model one should substitute a complex matrix $W$ by an anti-hermitian 
one. Let us write $W=i B$  
\begin{equation} 
D_3 = \left( \begin{array}{cc} 0 & iW \\ iW^\dagger & 0 
\end{array}\right) = 
\left( \begin{array}{cc} 0 & -B \\ B & 0 \end{array}\right) 
\label{D3} 
\end{equation} 
where $B=B^\dagger$ is from a GUE ensemble with the standard measure 
\begin{equation} 
\d\mu_3(B) = \d B e^{-\frac{N\Sigma^2}2 {\rm Tr} B^2} \ , 
\label{mu3} 
\end{equation} 
and where $\d B = \prod_{ii} \d B_{ii} \prod_{i<j} \d {\rm Re} B_{ij} \d
{\rm Im} B_{ij}$ is a flat measure in the set of hermitian
matrices. As for $D_4$ (\ref{D4}) the eigenvalues of the matrix $D_3$
are purely imaginary. They also come in pairs $\pm i \lambda$. The
microscopic eigenvalue density is however different. It is given by
the expression (\ref{rho3}), so it has neither a dip nor a wavy
structure. This shape just results from blowing up the central region
of the Wigner semicircle eigenvalue distribution of the matrix $B$
which is flat and has no hard edge at $\lambda=0$.  One expects that
the fermionic operator of QCD$_3$ should reproduce this structure in
the continuum limit. So we have to understand the eigenvalue density
observed for staggered fermions for quenched QCD$_3$ shown in figure
\ref{puzzle}.

\section*{Staggered fermions}

Staggered fermions discretization and its relation to continuum
physics in even dimensions was discussed in \cite{kmnp} and in odd
dimensions in \cite{bb}.  Let us shortly discuss how to calculate of the
spectrum of the staggered fermion operator in quenched simulations of
lattice QCD.  In our simulations we used the standard Wilson action to
generate gauge fields
\begin{equation} 
S_W = \beta \sum_x \sum_{\{\mu,\nu\}} 
\left(1 - \frac13 {\rm Re} {\rm Tr} 
U_\mu(x) U_\nu(x+\mu) U^\dagger_{\mu}(x+\mu+\nu) 
U^\dagger_{\nu}(x+\nu) \right) \ . 
\label{SW} 
\end{equation} 
where $U_\mu(x)$ is a $SU(3)$ matrix -- a dynamical variable
associated with a link going between two neighboring sites $x$ and
$x+\mu$.  The first sum runs over all lattice sites $x$, the second
one over all pairs of directions $1\le \mu < \nu \le D$, so $S_W$
collects contributions from all elementary plaquettes. In the quenched
approximation one generates gauge field configurations with the
probability measure $\prod dU e^{-S_W}$ which is independent of
fermionic degrees of freedom.  One uses these configurations to
compute quantum averages. In other words the influence of fermions on
gauge fields is neglected in this approximation.

The staggered fermion operator is defined as \cite{sf}
\begin{equation} 
D^{ij}_{xy} = \sum_{\mu=0}^d 
\eta_\mu(x)\left( U^{ij}_\mu(x) \delta_{y,x+\mu} - 
(U^\dagger)^{ij}_\mu(x-\mu) \delta_{y,x+\mu}\right) 
\label{staggered} 
\end{equation} 
where $ij$ are indices of the $SU(3)$ matrix and $\eta_\mu(x) =
(-1)^{x_1+\ldots+x_{\mu-1}}$ for $\mu=1,\ldots,D$.  One can also
define a parity operator $\epsilon(x) = (-1)^{x_1+\ldots + x_D}$, if the
lattice has even number of sites in each direction, that divides sites
into odd and even ones and introduces a chessboard structure. Each odd
site has even neighbors and vice versa, so the staggered fermion
matrix can be decomposed into blocks
\begin{equation} 
D = \left( \begin{array}{cc} 0 & 
D_{eo} \\ D_{oe} & 0 \end{array} \right) 
\end{equation} 
which are mutually anti-conjugated $D_{eo} = -D^{\dagger}_{oe}$ by
construction (\ref{staggered}). Writing $D_{oe} = i W$, one obtains a
matrix of the form (\ref{D3}).  Figure \ref{puzzle} tells us that the
$D_{eo}=iW$ behaves for $\beta=6.0$ as if $W$ belonged to the
universality class of Gaussian complex matrices. Since the even-odd
split $D_{eo} = -D_{oe}^\dagger$ is built into the discretization
scheme one can expect that the related microscopic universality is
robust unless something dramatic happens that changes the basic
symmetry or the nature of the randomness of the matrix.  An example of
such a mechanism may be a freeze-out of some degrees of freedom of the
matrix.

In fact, it has been shown that in the continuum limit the staggered fermions in three dimensions corresponds to four two component spinors with the action
\begin{equation}
S_F= \sum_{y,k} [\bar{u}(\sigma_{k}\otimes I)D_{\mu}u - 
\bar{d}(\sigma_{k}\otimes I)D_{\mu}d ]
\label{cont_theo}
\end{equation}
where $I$ is a two by two unit matrix. This can be equivalently
written as two copies of the four component fermions defined above.
Thus we expect that in the continuum limit, that is for
$\beta\rightarrow \infty$, the matrix $W$ becomes anti-hermitian, or
equivalently that the fluctuations of hermitian degrees of freedom are
suppressed in this limit.  In the next section we propose a random
matrix model describing such a gradual suppression. This model has one
parameter which interpolates between the regime where $W$ is a generic
random complex matrix (with both the hermitian and anti-hermitian
sectors) and the regime where it is an anti-hermitian one
$W=-W^{\dagger}$. As it is  shown below the model captures the
pattern of evolution with $\beta$ of the microscopic density observed
in the numerical QCD$_3$ simulations.

\section*{Freeze-out random matrix model}
The idea is to consider a random complex matrix $W$ which is a linear
combination $W=xA + iyB$ of two independent identically distributed
Gaussian hermitian matrices $A=A^\dagger$ and $B=B^\dagger$ with real
coefficient $x,y$. The matrices are generated with the probability
measure
\begin{equation} 
  d\mu(A,B) = \d A \d B e^{-\frac{N\Sigma^2}{2} {\rm Tr} A^2}  e^{-\frac{N\Sigma^2}{2}  {\rm Tr} B^2} \ , 
\label{muAB} 
\end{equation} 
where $\d A$ and $\d B$ are flat measures for hermitian matrices. 
To avoid redundancy with the width parameter $\Sigma$ it is convenient 
to restrict $x,y$ to $x^2+y^2=1$ or equivalently to parametrize $x=\cos(\alpha)$, $y=\sin(\alpha)$. We obtain a one-parameter 
family of matrices 
\begin{equation}
D_{\alpha} = 
\left( \begin{array}{cc} 0 & iW^\dagger \\ 
iW & 0 \end{array}\right) =
\left( \begin{array}{cc} 0 & i \cos(\alpha) A - \sin(\alpha) B  \\ 
i \cos(\alpha) A + \sin(\alpha) B & 0 \end{array}\right) \ , 
\label{Da}
\end{equation} 
with a mixing parameter $\alpha$ which interpolates between (\ref{D4}) and (\ref{D3}). For $\alpha=\pi/4$ this matrix is equivalent to (\ref{D4}) where $W=(A+iB)/\sqrt{2}$ (\ref{mu4}). The integration measure for $W$ can be
derived by changing variables in (\ref{muAB}) which gives:
\begin{equation}
d\mu(W) \sim \d W e^{ - N\Sigma^2{\rm Tr} WW^\dagger}
\end{equation}
This is a standard Girko-Ginibre ensemble \cite{g1,g2}. 
Eigenvalues of $W$ are uniformly
distributed in a disk of radius $1/\Sigma$ in complex plane centered around the origin. 
For arbitrary $\alpha$ it becomes an elliptic ensemble with a measure
\begin{equation}
\d\mu(W) \sim DW 
e^{ - \frac{N\Sigma^2}{(1-\tau^2)} {\rm Tr} \left(WW^\dagger - \frac{\tau}{2} \left( WW + W^\dagger W^\dagger\right)\right)}
\label{muW}
\end{equation}
where $\tau = \cos(2\alpha)$ \cite{scss}.  Eigenvalues are now
uniformly distributed in an elliptic disc with semi-axes of relative
length $(1-\tau)/(1+\tau)$. When $\alpha$ approaches $\pi/2$,
hermitian degrees of freedom are gradually suppressed. For
$\alpha=\pi/2$, $W$ becomes purely anti-hermitian (\ref{D3}) and the
ellipse reduces to a cut on the real axis. The evolution of the
eigenvalue spectrum of the matrix $D_\alpha$ for $\alpha$ from $\pi/4$
to $\pi/2$ (or equivalently $\tau$ from $0$ to $-1$) smoothly
interpolates between the limiting eigenvalue densities
$\rho_{4*}(\lambda)$ and $\rho_{3*}(\lambda)$ when $\alpha$.

It is instructive to begin the analysis of the spectrum of the matrix
$D_\alpha$ by considering the case $N=1$. Although it is a very
simplified version of the full model it allows one to understand the
nature of the hard edge of the spectrum and the mechanism of its
disappearance for $\alpha\rightarrow \pi/2$.  For $N=1$ the matrices
$A,B$ reduce to real numbers $a,b$ which are independent, identically
distributed normal random variables with the variance
$1/\Sigma^2$. The matrix (\ref{Da}) has only one pair of eigenvalues
\begin{equation} 
\lambda_\pm =  \pm i r = 
\pm i \sqrt{a^2\cos^2(\alpha)  + b^2\sin^2(\alpha)} \ . 
\label{r}
\end{equation} 
We are now interested in finding the probability distribution of the
random variable $r$ which is constructed from $a$ and $b$.  This
quantity has a geometrical interpretation. The vector
$\vec{r}=(\cos(\alpha)a,\sin(\alpha) b)$ can be viewed as a random
vector on a plane whose components are independent Gaussian variables
with the standard deviations $\Sigma^{-1}\cos(\alpha)$ and
$\Sigma^{-1} \sin{\alpha}$. The length of this vectors is equal to r
(\ref{r}). The probability distribution for this vector is an elliptic
Gaussian (lines of equal probability form ellipses) so it is easy to
derive the corresponding distribution of length $r$ of the vector
$\vec{r}$. For $\alpha=\pi/2$ the distribution reduces to a one
dimensional distribution concentrated on the $b$-axis. In this case
$r=|b|$ and thus the probability distribution of $r$ is a
half-Guassian distribution $p(r)=2\Sigma/\sqrt{2\pi}
e^{-\Sigma^2r^2/2}$. For $\alpha=\pi/4$ the distribution is a
spherically symmetric Gaussian one and the distribution of $r$ can be
easily calculated by changing to the polar coordinates and integrating
out the angle. It yields $p(r) = 2\Sigma^2 r e^{-\Sigma^2 r^2}$. For
$\alpha$ in-between the distribution can be expressed by the following
integral
\begin{equation} \label{pa}
p_\alpha(r) = \frac{\Sigma^2}{2\pi} \int_0^{2\pi} d\phi \ 
\frac{r}{\cos^2(\alpha) \cos^2(\phi) + \sin^2(\alpha) \sin^2(\phi)} 
\exp\left( -\frac{\Sigma^2 r^2}{2(\cos^2(\alpha) \cos^2(\phi) 
+ \sin^2(\alpha) \sin^2(\phi))} \right) 
\end{equation}
which interpolates between the two limits. The evolution of the
probability distribution with $\alpha$ is shown in
figure~\ref{model_N1}.  For small $r$ the integral behaves as
$p_\alpha(r) \sim 2\Sigma^2 r/\sin(2\alpha)$, so it goes linearly to
zero with a slope $2\Sigma^2/\sin(2\alpha)$.  The slope becomes
infinite for $\alpha \rightarrow \pi/2$. In this limit also the
position of the maximum goes to zero and the dip gets narrower. It
disappears completely only for $\alpha=\pi/2$.
\begin{figure} 
\begin{center}
\includegraphics[width=17cm]{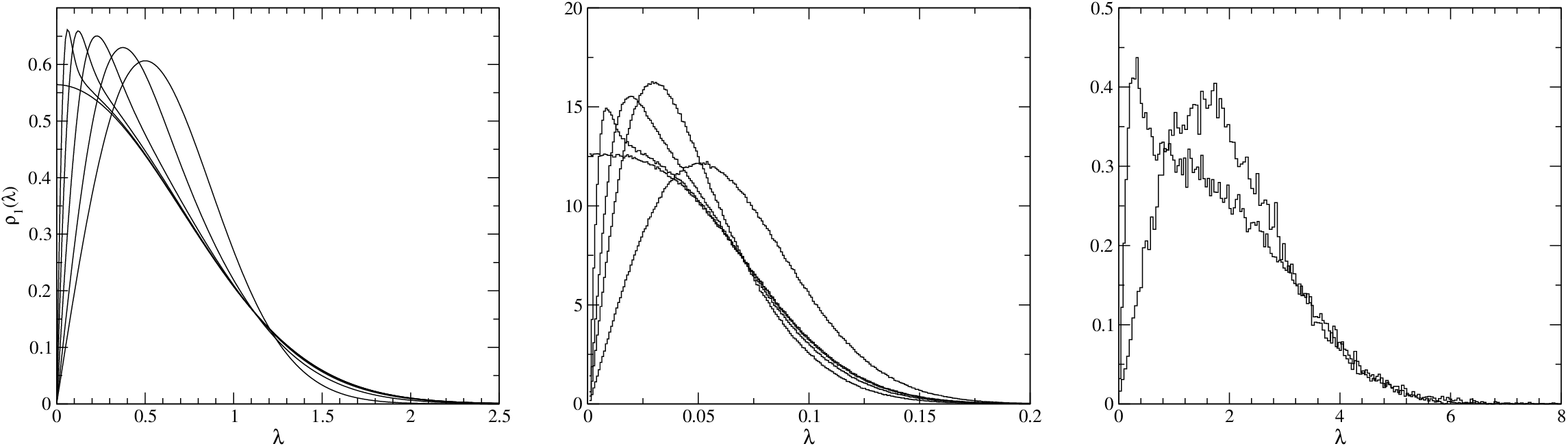}
\end{center}
\caption{\label{model_N1} 
The distribution $\rho_1(\lambda)$ of the lowest eigenvalue for the systems considered in this article.
(Left) The distribution \eqref{pa} $p_\alpha(r)$ for 
$\cot(\alpha)=1.0, \; 0.4, \; 0.2, \;  0.1, \;  0.5, \; 0.0$. We set
$\Sigma=1$. The maximum moves from right to left when $\alpha$ is decreased. 
The function $p_\alpha(r)$ goes linearly to zero 
$p_\alpha(r) \sim 2 r/\sin(2\alpha)$ for $r\rightarrow 0$ with 
a coefficient $2/\sin(2\alpha)$. The dip disappears only for $\sin(2\alpha)=0$.
(Middle) The distribution of the lowest eigenvalue obtained numerically 
for the matrix $D_\alpha$ in the matrix model for $N=20$ and for 
$\cot(\alpha)=0.0, 0.02, 0.05, 0.10, 1.0$. As before the maximum moves 
from right to left when $\alpha$ is increased from $\pi/4$ to $\pi/2$. 
The linear part of the slope at zero has a coefficient which 
increases when $\alpha$ approaches $\pi/2$.
(Right) QCD$_3$ for $\beta=12$ and $30$ on $24^3$ lattices. The eigenvalues are rescaled according to \eqref{micro}.  
} 
\end{figure} 
To summarize, we see that as long as $\alpha \ne \pi/2$ there 
is a repulsion from the hard edge at the origin. It can 
be attributed to the two-dimensional nature of the underlying 
distribution of the vector $(\cos(\alpha) a, \sin(\alpha) b)$ in the plane. 
The hard edge and the repulsion completely disappears only for 
$\sin(2\alpha)=0$ where the distribution becomes one-dimensional.
We expect the same mechanism of repulsion for the smallest eigenvalue
of the matrix $D_\alpha$ (\ref{Da}) also for $N>1$.
As long as both hermitian and anti-hermitian degrees of freedom are active
the lowest eigenvalues will be repelled from zero. Indeed we see in figure \ref{model_N1} in the middle that the evolution of the probability density for the lowest eigenvalue of the matrix $D_\alpha$ for $N=20$ very much resembles that shown in the left panel. For large $\alpha$ the 
distribution has a shape of the type $r e^{-r^2/2}$, 
while for small ones it is more like a semi-Gaussian shape with 
an additional narrow peak on top of it. This additional peak 
gets narrower when $\alpha \rightarrow \pi/2$ and it moves towards zero.
The same pattern is observed also for larger $N$. Finally, in the right chart
we show corresponding plots in QCD$_3$ data. 

The repulsion has also an influence on the second smallest eigenvalue, 
the third smallest one etc.  For $\alpha \rightarrow \pi/2$
the repulsion range becomes smaller and correspondingly 
the influence on higher eigenvalues becomes smaller too. 
The resulting microscopic eigenvalue distribution will be discussed in the next section where it will be compared to the corresponding histograms for quenched QCD$_3$ for different values of $\beta$.

\section*{Comparison of QCD$_\mathbf{3}$ to the random matrix model}

In figure \ref{QCDvsMM} we show histograms of eigenvalue distribution 
for QCD$_3$ for different values of $\beta=6.0; 12.0; 18.0$ for
$24^3$ lattice (lhs) and for the matrix model (\ref{D}) for different values 
of $\cot(\alpha)=1.0;0.1;0.05$ and $N=100$ (rhs). 
\begin{figure} 
\begin{center} 
\includegraphics[width=16cm]{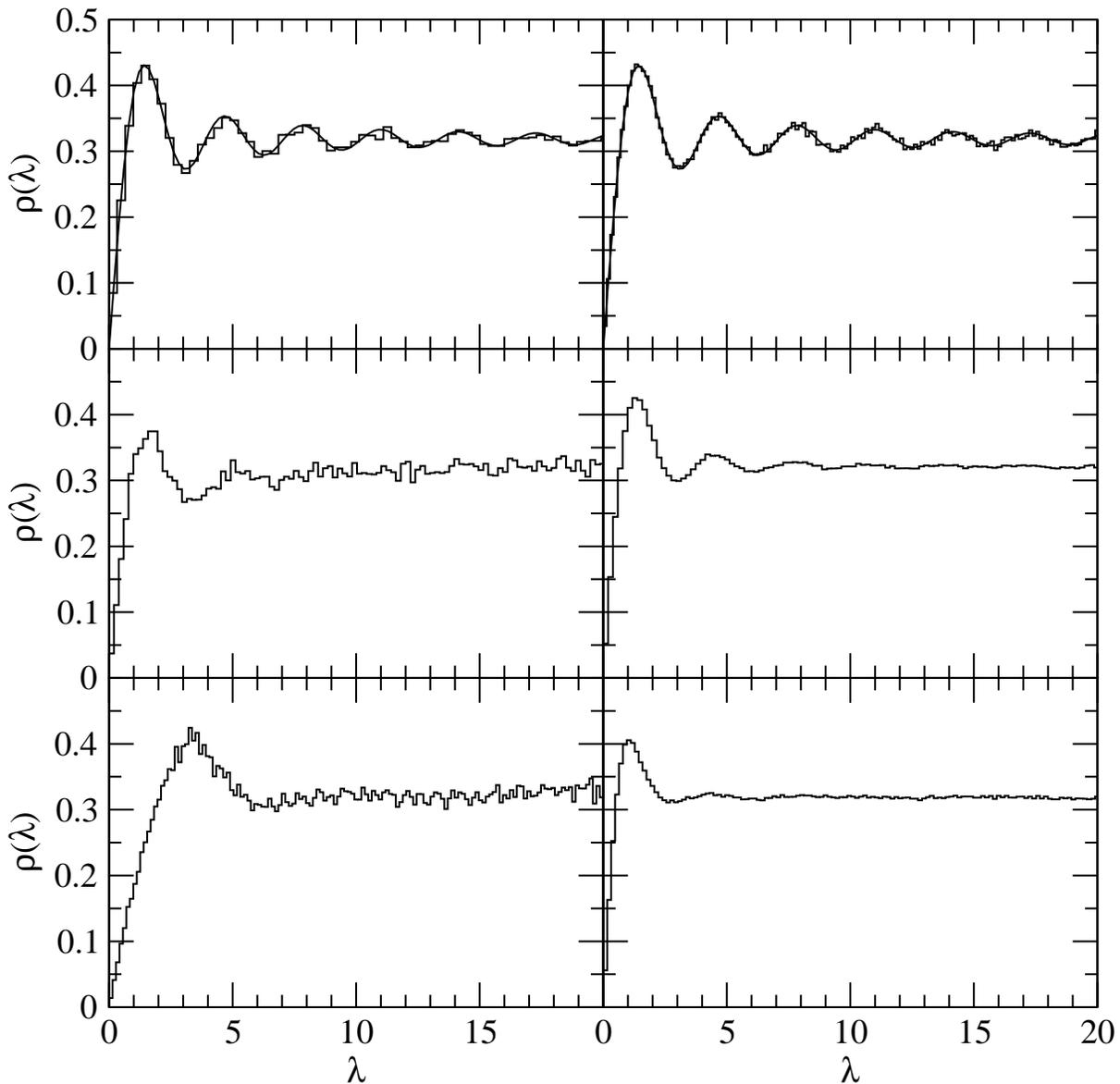}
\end{center} 
\caption{\label{QCDvsMM} 
Left: eigenvalue distribution for QCD$_3$ for 
$\beta=6.0; \; 12.0; 18.0$ for lattices 
$14^3$ , $24^3$, $24^3$ respectively. 
Right: eigenvalue distribution for $D_\alpha$ for the matrix 
model for $\cot(\alpha)=1.0; \; 0.1$ and $0.05$ for $50\times50$ matrices. 
}
\end{figure} 
When $\beta$ and $\alpha$ decrease the wavy 
structure gradually disappears in a very similar way in both cases
as one can see by comparing the figures. 
One should notice an anomaly in the plot in the bottom-left panel 
for $\beta=18.0$ where the main peak seems to be slightly wider 
than the corresponding peaks in other panels. The broadening is related
to an effect of eigenvalue pairing which appears when $\beta$ is increased. 
The pairing means that the distance between the first and the second
eigenvalue, the third and the fourth one, etc decreases when $\beta$ increases.
For $\beta=18.0$ the distribution of the first one and the second one are so
close to each other, as shown in figure \ref{split}, that they contribute to one broader peak in the eigenvalue density function
seen in the bottom-left plot in figure \ref{QCDvsMM}. 
\begin{figure} 
\begin{center} 
\includegraphics[width=17cm]{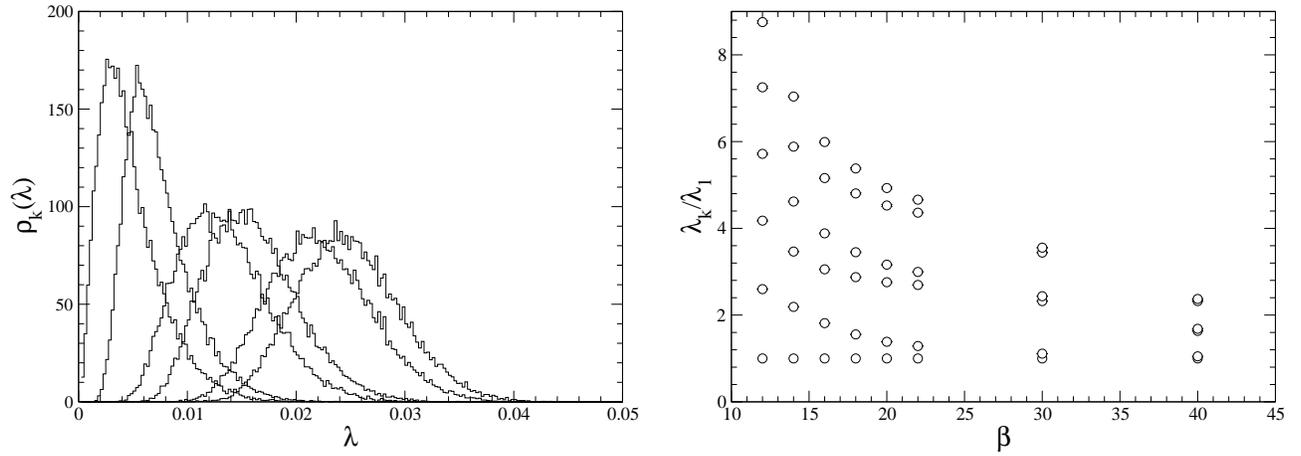}
\end{center} 
\caption{\label{split} 
(Left) Histograms of individual eigenvalues for QCD$_3$ with $\beta=18.0$ 
on $24^3$ lattice. (Right) Averages of the first six eigenvalues in
QCD$_3$ for $N=24^3$ as a function of $\beta$. All eigenvalues are normalized to the smallest eigenvalue for given $\beta$.} 
\end{figure}
For $\beta \rightarrow \infty$ the distance between neighboring eigenvalues tends to zero giving degenerate pairs of eigenvalues in the limit 
as one can see in the right chart in figure \ref{split}. This is in agreement with what one expects from the continuum theory  (\ref{cont_theo}) for which the Dirac operator indeed has doubly degenerate eigenvalues. 

\section*{Summary and discussion}

We showed that a simple two-matrix model reproduces basic
characteristic features observed in the eigenvalue spectra of the
Dirac operator in staggered QCD$_3$ data. The mixing parameter $\tau =
\cos 2\alpha$ interpolates between the regime where block matrices $W$
(\ref{muW}) in the Dirac matrix (\ref{Da}) are complex to the regime
where they are anti-hermitian. In the former case the microscopic
density is given by (\ref{rho4}) while in the latter one by
(\ref{rho3}). We observe a similar evolution in the QCD$_3$ data. The
shape of the probability distribution of the lowest eigenvalue evolves
in both the cases in a similar way too. The nature of finite size
effects manifesting as the appearance of a small narrow peak on top of
the distribution as well as the gradual disappearance of the hard core
close to the origin is similar in both the cases too. What remains to
show is how the microscopic density precisely interpolates between the
two limiting cases (\ref{rho4}) for $\alpha=\pi/4$ ($\tau=0$) and
(\ref{rho3}) for $\alpha=\pi/2$ ($\tau=-1$). We expect that a
continuous evolution of the microscopic density can observed in a
parameter $\widehat{\alpha} = \sqrt{N}(\pi/2 - \alpha)$ in a double
scaling limit, that is for $N \rightarrow \infty$, $\alpha \rightarrow
\pi/2$ and for finite $\widehat{\alpha}$. This limit corresponds to a
very asymmetric mixing of hermitian and anti-hermitian degrees of
freedom $W \approx (\widehat{\alpha}/\sqrt{N}) A + i B$.  The width of
fluctuations of hermitian degrees of freedom decreases with the system
size as $1/\sqrt{N}$ and the aspect ratio $\tau$ of the elliptic
ensemble (\ref{muW}) approaches minus one as $\tau = -1 +
4\widehat{\alpha}^2/N$. The evolution of the shape with
$\widehat{\alpha}$ can be found using for instance the supersymmetric
method \cite{VSS}.  Having done that one can try to relate the
parameter $\widehat{\alpha}$ to $\beta$ in QCD$_3$. We leave this
issue for future investigations.

\bigskip

\begin{center}
{\bf Acknowledgements} 
\end{center}
This work was supported by the EC-RTN Network ``ENRAGE'',
No.~MRTN-CT-2004-005616 and the Polish Ministry of Science Grant
No.~N~N202~229137 (2009-2012).


\begin{thebibliography}{99}
\bibitem{sv} E.V. Shuryak, J.J.M. Verbaarschot, Nucl.Phys. {\bf A560} 306 (1993);
\bibitem{vz4} J.J.M. Verbaarschot and I. Zahed, Phys. Rev. Lett. {\bf 70}, 3852 (1993);
\bibitem{sf} T. Banks, J.B. Kogut and Susskind, Phys. Rev. {\bf D 13} 1043 (1976); L. Susskind, Phys. Rev. {\bf D 16} 3031 (1977);
\bibitem{dhk} P.H. Damgaard, U.M. Heller and A. Krasnitz, Phys. Lett. {\bf B445}, 366 (1999);

\bibitem{follana} E.~Follana, A.~Hart, C.T.H.~Davies, Q.~Mason,
  Phys.~Rev.~D{\bf72} (2005) 054501.

\bibitem{vz3} J.J.M. Verbaarschot and I. Zahed, Phys. Rev. Lett. {\bf 73}, 2288 (1994);


\bibitem{dhkm} P.H. Damgaard, U.M. Heller, A. Krasnitz and T. Madsen
Phys.Lett. B440 (1998) 129;
\bibitem{vw} J.J.M. Verbaarschot and T. Wettig, Ann. Rev. Nucl. Part. Sci. {\bf 50} 343 (2000);
\bibitem{v} J.J.M. Verbaarschot,
{\em QCD, Chiral Random Matrix Theory and Integrability},
Les Houches lectures notes, arXiv:hep-th/0502029;
\bibitem{p} R.D. Pisarski, Phys. Rev. {\bf D29}, 2423 (1984);


\bibitem{kmnp} H. Kluberg-Stern, A. Morel, O. Napoly and B. Petersson,
Nucl. Phys. {\bf B 220} [FS8] 447 (1983);

\bibitem{bb} C. Burden and A.N. Burkitt, Europhys. Lett. 3 (5) 545 (1987);



\bibitem{g1} J.Ginibre, J. Math. Phys. {\bf 6}, 440 (1965)
\bibitem{g2} V. L. Girko, {\it Spectral theory of random matrices}, in Russian (Nauka, Moscow 1988), and references therein.
\bibitem{scss} H.-J. Sommers, A. Crisanti, H. Sompolinsky, and Y. Stein, Phys. Rev. Lett. {\bf 60}, 1895 (1988);
\bibitem{VSS} J.J.M. Verbaarschot, 
{\em The Supersymmetric Method in Random Matrix Theory and Applications to QCD}, 
Lectures given at the 2004 ELAF Summer School in Mexico City
arXiv:hep-th/0410211;
\end{thebibliography}
\end{document}